\documentclass[%
 prl,preprint,groupedaddress,
 amsmath,amssymb,
 aps,
]{revtex4-1}

\usepackage{graphicx}
\usepackage{dcolumn}
\usepackage{bm}

\begin{document}


\title{Total Width of 125 GeV Higgs Boson}

\author{Vernon Barger}
 \email{barger@pheno.wisc.edu}
\affiliation{
Department of Physics, University of Wisconsin, Madison, WI 53706, USA
}%

\author{Muneyuki Ishida}
 \altaffiliation[ ]{Department of Physics, University of Wisconsin-Madison. A visitor until March 2012.}
 \email{mishida@wisc.edu}
\affiliation{%
Department of Physics, Meisei University, Hino, Tokyo 191-8506, Japan
}%


\author{Wai-Yee Keung}
 \email{keung@uic.edu}
\affiliation{
Department of Physics, University of Illinois at Chicago, IL 60607, USA
}%


\date{\today}

\begin{abstract}
By using the LHC and Tevatron measurements of the cross sections to various decay channels 
relative to the standard model Higgs boson, the total width of the putative 125 GeV Higgs boson
is determined as 6.1$\stackrel{+7.7}{\scriptstyle -2.9}$~MeV. 
We describe a way to estimate the branching fraction for Higgs decay to 
dark matter.
We also discuss a No-Go theorem for the $\gamma\gamma$ signal of the Higgs boson at the LHC. 
\end{abstract}

\maketitle

The total width of the 125 GeV Higgs-boson signal is of intrinsic interest, but
it is generally very difficult to determine the total width $\Gamma_{\rm tot}$ of a narrow resonance
like the Higgs boson. 
Moreover,
the determination of this quantity is also an important test of the Higgs mechanism of the Standard Model (SM).
A sizable deviation from the SM prediction would directly indicate new physics.
The Higgs width can be measured at a $\gamma$-$\gamma$ collider\cite{GH} or 
a $\mu^+\mu^-$ collider\cite{BBGH} through its line shape, and such facilities are under
consideration.

We will present a simple method to determine the total width $\Gamma_{h^0}^{\rm tot}$ 
of the 125 GeV Higgs signal $h^0$ by using LHC and Tevatron measurements 
with the SM Higgs boson $h_{\rm SM}^0$ as a benchmark.
We will apply this method to the data of the putative Higgs-boson signal with mass 125~GeV.
Pre-LHC studies\cite{Zeppen,Belyaev,Duhrssen,Lafaye} were made of a similar ilk. 

\noindent\underline{\it Outline of the method}\ \ \ 
The $h^0$ total width is given by the sum of partial widths that can be normalized to the
$h^0_{\rm SM}$ partial widths.
\begin{eqnarray}
\Gamma_{h^0}^{\rm tot} &=& \sum_{A\bar A}\Gamma_{h^0\rightarrow A\bar A} 
 = \sum_{A\bar A} \gamma_{AA} \Gamma_{h^0_{\rm SM}\rightarrow A\bar A},\ \ \ 
\gamma_{AA} = \frac{\Gamma_{h^0\rightarrow A\bar A}}{\Gamma_{h^0_{\rm SM}\rightarrow A\bar A}}\ .
\label{eq1}
\end{eqnarray}
For the 125 GeV Higgs, we consider the channels 
$A\bar A=b\bar b,\tau\tau,gg,WW^*,ZZ^*,c\bar c,\gamma\gamma,Z\gamma$.
$\gamma_{AA}$ is the ratio of the $h^0$ partial width of the $A\bar A$ channel to that of $h^0_{\rm SM}$. 
The cross sections of a given channel relative to the $h^0_{\rm SM}$ expectation is given by
\begin{eqnarray} 
XA &\equiv &\frac{\sigma (X\bar X\rightarrow h^0 \rightarrow A\bar A)}{\sigma (X\bar X\rightarrow h_{\rm SM}^0 \rightarrow A\bar A)}
 = \frac{\gamma_{XX}\gamma_{AA}}{\Gamma_{h^0}^{\rm tot}/\Gamma_{h^0_{\rm SM}}^{\rm tot}}
\label{eq2}
\end{eqnarray}
where $X$ is the initial parton in the proton participating in the fusion process. 
Then, we can obtain $\Gamma_{h^0}^{\rm tot}$ via measurements of 
the ratios of Eq.~(\ref{eq2}) following Eq.~(\ref{eq1}).
Equation (\ref{eq2}) is derived from the proportionality 
of $\sigma (X\bar X\rightarrow h^0\rightarrow A\bar A)$ to the corresponding decay width 
$\Gamma_{h^0\rightarrow X\bar X}$ 
and the branching fraction $BF(h^0\rightarrow A\bar A)$, that is,
$
\sigma (X\bar X\rightarrow h^0 \rightarrow A\bar A) \propto  
\Gamma_{h^0\rightarrow X\bar X}\cdot BF(h^0\rightarrow A\bar A)\ 
$.

In Eq.~(\ref{eq1}),
$\Gamma_{h^0}^{\rm tot}/\Gamma_{h^0_{\rm SM}}^{\rm tot}$ is represented by
\begin{eqnarray}
\Gamma_{h^0}^{\rm tot}/\Gamma_{h^0_{\rm SM}}^{\rm tot} 
 &\equiv& R = 0.58\gamma_{bb}+0.06\gamma_{\tau\tau}+0.24\gamma_{VV}+0.09\gamma_{gg}+0.03\gamma_{cc},
\label{eq3}
\end{eqnarray} 
where we use the $BF$ of $h^0_{\rm SM}$ in Table~\ref{tab1}, 
extracted from Ref.\cite{Heinemeyer}
and assume  $\gamma_{WW^*}=\gamma_{ZZ^*}(\equiv\gamma_{VV})$
as is the case for spontaneous symmetry breaking via the $SU(2)_L$ Higgs 
doublet.\cite{note1}
$\gamma_{cc}$ can be approximated by unity in Eq.~(\ref{eq3}) since $\gamma_{cc}$ is a 
subleading contribution.

\begin{table}[htb]
\begin{tabular}{c|cccccccc}
Channel & $b\bar b$ & $\tau^-\tau^+$ & $WW^*$ & $ZZ^*$ & $gg$  & $c\bar c$ & $\gamma\gamma$ & $Z\gamma$\\
\hline
Br(\%)  & 57.7 & 6.32 & 21.5 & 2.64 & 8.57 & 2.91 & 0.228 & 0.154\\ 
\hline
\end{tabular}
\caption{Branching fractions (BF) of the SM Higgs boson with mass 125 GeV as predicted in ref.\cite{Heinemeyer}. 
The total Higgs width is $\Gamma_{h^0_{\rm SM}}^{\rm tot}=4.07$~MeV with an uncertainty of $\pm 4$\%.}
\label{tab1}
\end{table} 

\noindent\underline{\it Illustrations of width determination}\ \ \ 
The 5 $\gamma$-parameters, $(gg,b\bar b,\tau^+\tau^-,VV$ and $\gamma\gamma )$, can be 
determined by LHC and Tevatron measurements of the corresponding ratios in Eq.~(\ref{eq2}). 
Then the value of $\Gamma_{h^0}^{\rm tot}$ is determined by
\begin{eqnarray}
\Gamma_{h^0}^{\rm tot} &=& \Gamma_{h^0_{\rm SM}}^{\rm tot}\cdot 
 (0.58\gamma_{bb}+0.06\gamma_{\tau\tau}+0.24\gamma_{VV}+0.09\gamma_{gg}+0.03)
\label{eq4}
\end{eqnarray}
with $\Gamma_{h^0_{\rm SM}}^{\rm tot}=4.07$~MeV\cite{Heinemeyer}.
The small $\gamma\gamma$ and $Z\gamma$ contributions can be neglected here.

The experimental values of the ratios of Eq.~(\ref{eq2})
at $m_{h^0}=125$~GeV reported by CMS\cite{CMS1,CMS} 
and by ATLAS\cite{ATLAS} are given in Table~\ref{tab2}, along with
the ratio for the $b\bar b$ channel inferred from the latest Tevatron data\cite{Tevatron}.
A $\chi^2$ fit gives the estimated values of the $\gamma_{AA}$ parameters in Table~\ref{tab3}. 

\begin{table}
\begin{tabular}{c|c|ccc}
$\sigma/\sigma_{\rm SM}$ &  & CMS\cite{CMS,CMS1} & ATLAS\cite{ATLAS} & Tevatron\cite{Tevatron} \\
\hline
$q\bar q\rightarrow Vb\bar b$ & $Vb=\gamma_{VV}\cdot\gamma_{bb}/R$ & 1.2$\stackrel{+2.1}{\scriptstyle -1.9}$
  & -0.8$\stackrel{+1.8}{\scriptstyle -1.7}$ & 2.0$\pm 0.7$\\
$gg\rightarrow \tau^- \tau^+$ & $g\tau=\gamma_{gg}\cdot\gamma_{\tau\tau}/R$
  & 0.63$\stackrel{+1.00}{\scriptstyle -1.28}$ & 0.0$\pm$1.7 & \\ 
$gg\rightarrow \gamma\gamma$ & $g\gamma=\gamma_{gg}\cdot\gamma_{\gamma\gamma}/R$
  & 1.62$\pm 0.68$ & 1.6$\stackrel{+0.8}{\scriptstyle -0.7}$ &  \\ 
$gg\rightarrow WW^*$ & $gW=\gamma_{gg}\cdot\gamma_{VV}/R$
  & 0.40$\pm 0.55$ &  0.20$\pm 0.62$ & 0.0$\stackrel{+1.0}{\scriptstyle -0.0}$  \\ 
$gg\rightarrow ZZ^*$ & $gZ=\gamma_{gg}\cdot\gamma_{VV}/R$
  & 0.58$\stackrel{+0.94}{\scriptstyle -0.58}$ & 1.4$\stackrel{+1.3}{\scriptstyle -0.8}$ & \\
$VV\rightarrow \gamma\gamma$ & $V\gamma=\gamma_{VV}\cdot\gamma_{\gamma\gamma}/R$ 
 & 3.8$\stackrel{+2.4}{\scriptstyle -1.8}$\cite{CMS1} \\ 
$q\bar q\rightarrow VA\bar A$ & $qA=\gamma_{VV}\gamma_{AA}/R$ & \\
\hline
\end{tabular}
\caption{
$XA\equiv \sigma (X\bar X\rightarrow h^0 \rightarrow A\bar A)/\sigma (X\bar X\rightarrow h_{\rm SM}^0 \rightarrow A\bar A)$:
The observed Higgs-signal cross section at the LHC from various processes relative to the standard model Higgs at $m_{h^0}=125$~GeV 
are given by CMS\cite{CMS} 
and by ATLAS\cite{ATLAS}.
$VV\rightarrow \gamma\gamma$ is determined by CMS from di-jet diphoton events\cite{CMS1}.
The $b\bar b$ signal of the second row is inferred from the recent Tevatron data\cite{Tevatron}.
$R$ is the $h^0$ total width relative to that of $h^0_{\rm SM}$ with the same mass. See Eq.~(\ref{eq3}).
}
\label{tab2}
\end{table} 

\begin{table}
\begin{tabular}{c|cccccc}
$A\bar A$ & $b\bar b$ & $\tau^- \tau^+ $ & $WW^*$ & $ZZ^*$ & $gg$ & $\gamma\gamma$ \\
\hline
$\gamma_{AA}$ & 1.8$\stackrel{+3.1}{\scriptstyle -1.1}$ & 1.1$\stackrel{+3.8}{\scriptstyle -2.7}$ 
 & $1.34\stackrel{+0.57}{\scriptstyle -0.45}$ & 1.34$\stackrel{+0.57}{\scriptstyle -0.45}$
 & 0.57$\stackrel{+0.48}{\scriptstyle -0.25}$ & 4.3$\stackrel{+5.2}{\scriptstyle -1.8}$\\
\hline
$BF(h^0\rightarrow A\bar A)$(\%) & 68.7$\stackrel{+14.9}{\scriptstyle -17.1}$
 & 4.5$\stackrel{+16.0}{\scriptstyle -11.3}$
 & 19.1$\stackrel{+19.1}{\scriptstyle -12.4}$ & 2.3$\stackrel{+2.3}{\scriptstyle -1.4}$
 & 3.2$\stackrel{+3.9}{\scriptstyle -2.2}$ & 0.65$\stackrel{+0.98}{\scriptstyle -0.45}$\\
\hline
$\Gamma_{h^0\rightarrow A\bar A}$(MeV) & 4.2$\stackrel{+7.3}{\scriptstyle -2.6}$
 & 0.3$\stackrel{+1.0}{\scriptstyle -0.7}$
 & 1.2$\stackrel{+0.5}{\scriptstyle -0.4}$ & 0.14$\stackrel{+0.06}{\scriptstyle -0.04}$
 & 0.20$\stackrel{+0.16}{\scriptstyle -0.09}$ & 0.04$\stackrel{+0.05}{\scriptstyle -0.02}$\\
\hline
\end{tabular}
\caption{ $\gamma_{AA}$ obtained by the fit to the data in Table~\ref{tab2}. 
One-sigma statistical uncertainties are given.
Partial widths of the 125 GeV Higgs 
$\Gamma_{h^0\rightarrow A\bar A}$ and the $BF$ are also given. 
The total width is estimated to be 
$\Gamma_{h^0}^{\rm tot}=6.1\stackrel{+7.7}{\scriptstyle -2.9}$~MeV.
The errors of $\Gamma_{h^0}^{\rm tot}$ and of $BF(h^0\rightarrow b\bar b)$ 
correspond to the one standard deviation of $\gamma_{bb}$. 
The $BF$ errors for the other channels 
are estimated by treating $\gamma_{AA}$ and $\Gamma_{h^0}^{\rm tot}$ as independent quantities.
In the SM all the $\gamma_{AA}$ are unity and the total width is 
$\Gamma_{h^0_{\rm SM}}^{\rm tot}=4.07$~MeV\cite{Heinemeyer}.
}
\label{tab3}
\end{table} 

Because of the strong correlations between $\gamma_{bb,\tau\tau}$ and $\gamma_{\gamma\gamma}$,
loose upper limits are obtained for these quantities from the present data.
We obtain the value 
$\Gamma_{h^0}^{\rm tot}=6.1\stackrel{+7.7}{\scriptstyle -2.9}$~MeV. 
However, the determination will be much improved (see e.g. Refs.\cite{Zeppen,Duhrssen}) as the data increase.
The 2012 LHC run is expected to accumulate an integrated luminosity 
15~fb$^{-1}$ per experiment at 8~TeV and those data can be combined with the data from the 
 5~fb$^{-1}$ at 7~TeV. 
If all the LHC uncertainties become half the present ones and the central values remain the same, 
the value of $\Gamma_{h^0}^{\rm tot}$ becomes 
$\Gamma_{h^0}^{\rm tot}=3.4\stackrel{+2.3}{\scriptstyle -1.5}$~MeV.

\noindent\underline{\it $\gamma\gamma$ Enhancement}\ \ \ \ 
The $\gamma\gamma$ cross section seems to be enhanced compared with $h^0_{\rm SM}$,
although this could be an upward statistical fluctuation.
In the SM it is given theoretically by the triangle loop diagrams of the $W$-boson and top quark.
If a new heavy fermion and$/$or a heavy scalar couple to the SM Higgs, and their masses are 
generated by the Higgs mechanism,
the $\gamma_{\gamma\gamma}$ and $\gamma_{gg}$ are given (see e.g. Ref.\cite{Rll}) by
\begin{eqnarray} 
\gamma_{\gamma\gamma}
 &=& \left( \frac{\frac{7}{4}1.19-\frac{4}{9}1.03-\frac{N_c}{3}Q_f^2
   -\frac{N_c}{12}Q_S^2}{\frac{7}{4}1.19-\frac{4}{9}1.03} \right)^2,\ \ 
 \gamma_{gg} = \left( \frac{-\frac{1}{6}1.03-\frac{C_f}{3}
   -f_S\frac{C_S}{12}}{-\frac{1}{6}1.03} \right)^2,
\label{eq5}
\end{eqnarray}
where the 1st(2nd) term in the numerator or denominator in $\gamma_{\gamma\gamma}$ represent
the $W$-boson(top quark) loop and 1.19(1.03) is the correction from the finite $W(t)$ mass.
The first terms in the denominator and the numerator in $\gamma_{gg}$ are from the top quark loop.
Here we have assumed the $h^0$ couplings to $W$ and $t$ are the same as those of $h^0_{\rm SM}$.
The masses of the new particles are assumed to be sufficiently heavy
that the mass corrections can be neglected.
The numerators and denominators are normalized in Eq.~(\ref{eq5}) to a fermion contribution. 
$Q_{f,S}$ is the electric charge of a new fermion(scalar). $N_c$ is the color degree of freedom 
of a new particle in the loop.
$C_{f,S}$ is the quadratic color Casimir factor of the new fermion(scalar). 
It is $1/2(3)$ in the fundamental(adjoint) representation; $f_S=1(1/2)$ for a complex(real) scalar.
It is an important conclusion that a new fermion or scalar contribution, 
if it does not have large $N_c$, works to decrease $\gamma_{\gamma\gamma}$.\cite{note3}
For example, 
in the 4th generation model, $(\gamma_{\gamma\gamma},\gamma_{gg})=(0.21,8.7)$. 
The large $\gamma_{gg}$ of the 4th generation leads to
$R(=\Gamma_{h^0}^{\rm tot}/\Gamma_{h^0_{\rm SM}}^{\rm tot})=1.66$, and correspondingly
the $WW^*,ZZ^*,b\bar b,\tau^-\tau^+$ channels from gluon fusion are enhanced by $\gamma_{gg}/R=5.2$ ; 
$\gamma\gamma$ via gluon fusion is $\gamma_{gg}\gamma_{\gamma\gamma}/R=1.1$, 
almost the same as the SM, while $\gamma\gamma$ by vector-boson fusion is strongly suppressed, 
$\gamma_{\gamma\gamma}/R=0.12$.   
If the $\gamma\gamma$ enhancement in the di-photon di-jet events\cite{CMS1} is mainly from
$VV$ fusion and the measured value is confirmed,
the fourth-generation model will be excluded mass-independently. 
See also ref.\cite{Ishikawa}.    
Similarly, if the enhancement of $VV\rightarrow h^0\rightarrow \gamma\gamma$ is confirmed, the interpretation 
of the 125~GeV Higgs signal as the dilaton or radion\cite{dr,Hooman,Kingman} will be discarded, 
since the vector-boson fusion to diphoton cross section is strongly suppressed in these models 
compared with the SM Higgs.

\begin{figure}[htb]
\begin{center}
\resizebox{0.5\textwidth}{!}{
  \includegraphics{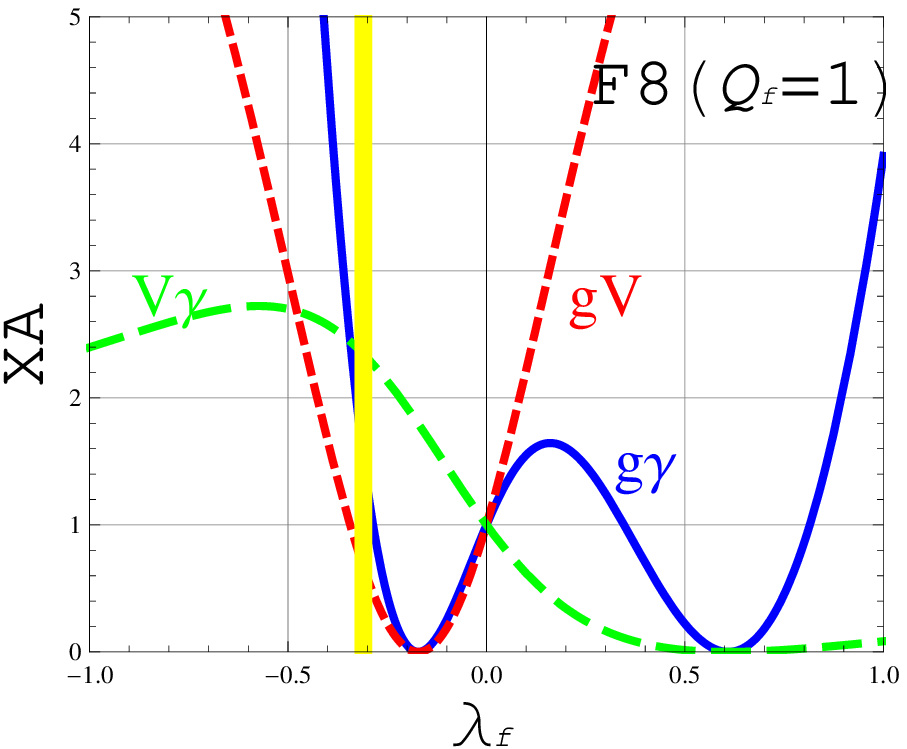}
}\\
\resizebox{0.5\textwidth}{!}{
  \includegraphics{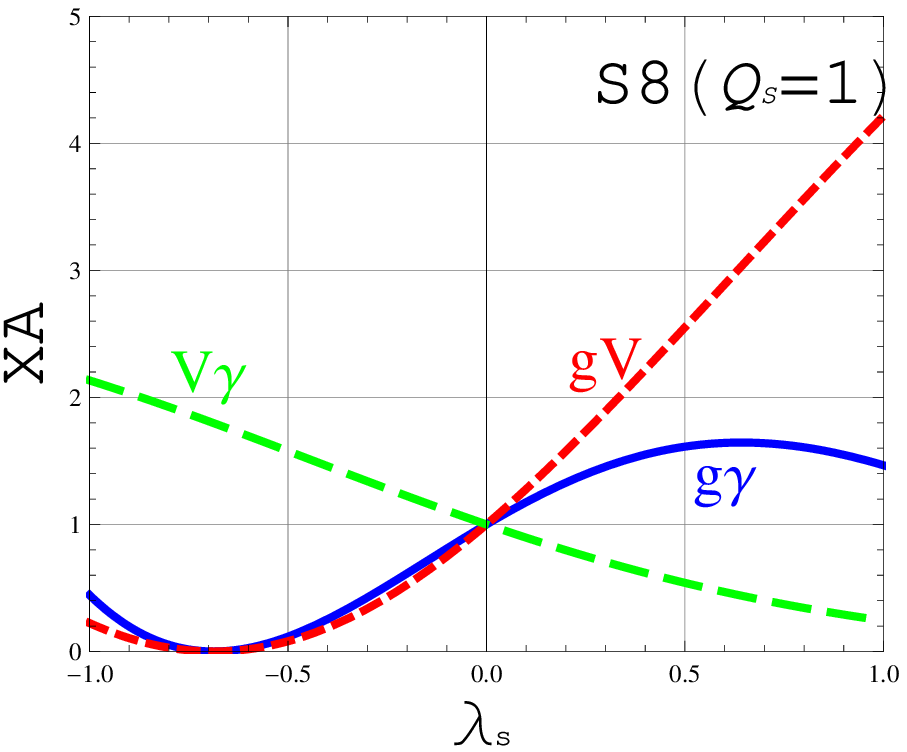}
}
\end{center}
\caption{$\lambda_{f,S}$ dependence of the cross sections of various processes relative to the SM Higgs, for the case of a color-octet fermion (leptogluon:$F8$) with charge $Q_s=1$ and  color-octet scalar ($S8$) with $Q_s=1$:
$XA \equiv \sigma(X\bar X\rightarrow h^0\rightarrow A\bar A)
/\sigma(X\bar X\rightarrow h^0_{\rm SM}\rightarrow A\bar A)$.
We consider three quantities $XA=g\gamma,gV,$ and $V\gamma$, corresponding to 
$gg\rightarrow \gamma\gamma$(solid blue),
$gg\rightarrow VV(WW^*$ or $ZZ^*)$(short-dashed red), 
and $VV\rightarrow \gamma\gamma$(long-dashed green).
$\lambda_{f,S}$ is the Higgs coupling normalized by the  
Yukawa couplings giving the masses by the Higgs mechanism. See Eq.~(\ref{eq6}) for definition. 
The yellow vertical band in the top panel is preferred by the present data suggesting $\gamma\gamma$ enhancement. 
}
\label{fig1}
\end{figure}

The loop contributions of a new scalar or a new fermion are proportional to 
dimensionless factors $\lambda_{f,S}$ 
\begin{eqnarray}
\lambda_f &=& \frac{{\cal Y}_f\ v}{m_f},\ \ \ \lambda_S = \frac{{\cal Y}_{h^0SS}\ v}{2m_S^2}
\label{eq6}
\end{eqnarray}
where 
${\cal Y}_f({\cal Y}_{h^0SS})$ is the Yukawa coupling of the new fermion (scalar)
and the $v$ is the Higgs VEV $v\simeq 246$~GeV. 
$\lambda_{f,S}=1$ corresponds to the 
the case that the fermion(scalar) mass is generated by the Higgs mechanism.
For a heavy particle with no Higgs mechanism for its mass generation, 
$\lambda_{f,S}\ll 1$ and $\gamma_{gg}\simeq \gamma_{\gamma\gamma}\simeq 1$, so the 
$\gamma\gamma$ cross section becomes the same as the SM Higgs.
To obtain a large enhancement, a $m_{f,S}$ smaller than $v$ is necessary
or alternatively the color factor of the new particle is large.  

The cross section ratios of various processes relative to the SM Higgs are plotted 
versus $\lambda_{f,S}$ in the cases of
color-octet fermion (denoted as $F8$, also called leptogluon\cite{leptogluon})
and color-octet scalar ($S8$)
in Fig.~\ref{fig1}. 

\begin{figure}[htb]
\begin{center}
\resizebox{0.5\textwidth}{!}{
  \includegraphics{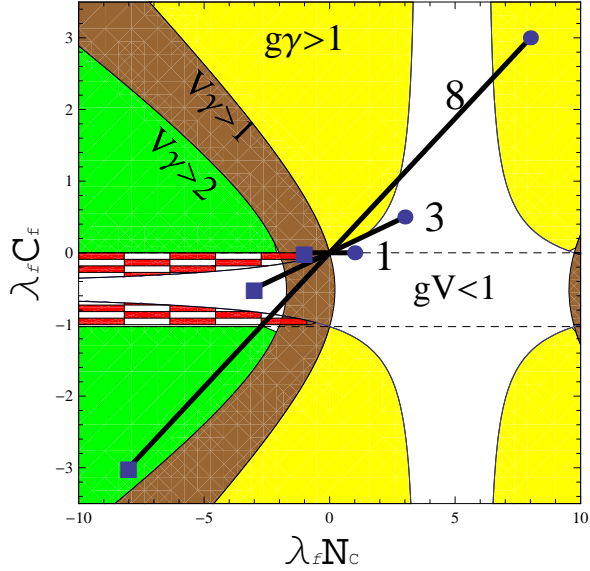}
}
\end{center}
\caption{Regions of $\gamma\gamma$ enhancement in $\lambda_fN_cQ_f^2,\ \lambda_fC_f$ plane in $Q_f=1$ case.
The $g\gamma>1$ region is divided into 4 colored regions:
($V\gamma<1$;$1<V\gamma<2$;$2<V\gamma$) are (yellow;brown;green), respectively.
Red meshed region, which is preferred by the present experimental data, corresponds to 
$V\gamma>1$ and $gV<1$, where the latter is between the two horizontal lines, $\lambda_fC_f=0,-1.03$.
Color-octet fermion(leptogluon),  Color-triplet fermion, and Color-singlet fermion 
are shown by solid lines
with the end points corresponding to $\lambda_f=-1$(square) and $\lambda_f=1$(circle).
In $Q_f\neq 1$ case,
the $x$-coordinates scale with $Q_f^2$.
$\lambda_f=1$ corresponds to the case of its mass generated by Higgs mechanism.
A color-octet fermion with $Q_f=1$ is consistent with the red meshed region 
at $\lambda_f\simeq -0.31$.
For a new scalar, the lengths of the theory lines should be scaled by $1/4$; 
thus, a scalar octet has no overlap with the preferred red meshed region. 
}
\label{fig2}
\end{figure}

The $S8$\cite{Schumann} is an interesting possibility.
It was discussed in the context of Higgs underproduction\cite{Dob,Fan} at LHC 
for the circumstance that this new scalar has light mass and the Higgs boson
has sizable branching fraction to this scalar channel. 
If the mass of this color-octet scalar is generated by the Higgs mechanism, 
following Eq.~(\ref{eq5}), by using $N_c=8(C_S=3)$,
the $\gamma$-values of a $Q_s=1$ charged scalar are
 $(\gamma_{\gamma\gamma},\gamma_{gg})=(0.35,6.0)$. 
For a new scalar without a  Higgs origin for its mass generation, 
the sign of the coupling $\lambda$ is arbitrary.
In the $S8$ case of Fig.~\ref{fig1} both enhancement factors, 
$g\gamma$ and $V\gamma$, are less than $\sim 2$. 
Similar results are also obtained for a color-triplet scalar($S3$ : leptoquark) 
and a color-triplet fermion ($F3$).
However,
the present data seem to suggest $WW^*$ suppression and $\gamma\gamma$ enhancement
in $gg$-fusion and $VV$-fusion. 
This tendency is not reproduced by $S8$ but may be realized with $F8$ as can be seen in Fig.~\ref{fig2},
where
the preferred regions of parameters, $\lambda_fN_cQ_f^2$ and $\lambda_f C_f$,  
by present data are shown.
In the case $F8$(leptogluon),
the trends of the present data can be reproduced with $\lambda_f \approx -0.31$
as shown by the yellow vertical band of $F8$ in Fig.~\ref{fig1}, for which
$(g\gamma,gV,V\gamma) = (1.52, 0.67, 2.35)$.
If the Yukawa coupling of the leptogluon is the same as top quark, but the sign of $\lambda$ is reversed,
its mass is estimated to be $m_{F8}=m_t/0.31\simeq 500$~GeV.  
The corresponding ${Z\gamma}$ partial decay width is almost the same as the SM Higgs:
 $\gamma_{Z\gamma}=1.04$. Then, a $Z\gamma$ cross section ratio via gluon fusion
$0.69\pm 0.09$ 
is predicted, which is almost the same as $gV$.

More generally, Fig.~\ref{fig2} is a vector space that can be used in identifying
new particle contributions from experimental measurements of the $XA$.

Solutions satisfying $g\gamma >1$, $V\gamma>2$, and $gV<1$ 
for lower-dimensional representations of $SU(3)_c$
are summarized in Table~\ref{tab4}.

\begin{table}
\begin{tabular}{l|ll||l|ll}
     &  $Q_{f}$ & $\lambda_{f}$ & & $Q_s$ & $\lambda_s$\\
\hline
$F1$ & $5/3\ (2)$ & -0.9(-0.7) & & & \\
$F3$ & $5/3\ (2)$ & -0.2(-0.3) & $S3$ &  $5/3\ (2)$         &  -1.0 (-0.9)\\
$F6$ & $1(5/3)$ & -0.38(-0.32) & & & \\
$F8$ & $1(4/3)$ & -0.31(-0.3) & & & \\
$F10$ & $4/3\ (2)$ & -0.13(-0.12) & $S10$ & $4/3\ (2)$ & -0.50(-0.45)\\
$F27$ & $5/3\ (2)$ & -0.034(-0.032) & $S27$ & $5/3\ (2)$  &  -0.14(-0.13)\\
\hline
\end{tabular}
\caption{Solution satisfying the $\gamma\gamma$ enhancement
for $SU(3)$ representations with dimensions$\le 27$.
$F3,S10$ represent the color-triplet fermion, color-decouplet scalar, for example.
The typical values of $\lambda$ satisfying $g\gamma>1,V\gamma>2,$  and $gV<1$ are given. 
}
\label{tab4}
\end{table}
It is very difficult to obtain $V\gamma>2$ and $g\gamma>1$.
This constitutes a NO GO theorem.
Only the $F8$ (and $F6$) are possible if we limit the charge of the new particle $Q_f\le 1$.
For still higher dimensional color representations, the theory lines 
do not overlap with the preferred region
for the $Q\le 1$ case as can be deduced from Fig.~\ref{fig2}.

The possibility of light stop and light stau in the MSSM are discussed 
in ref.\cite{falkowski,maiani,carena}\cite{note4}.
The effect of the stop loop is suppressed compared with top-quark loop 
because stop quark is scalar, 
and the chargino contribution is suppressed by the absence of color.
The stau effect is suppressed by both. 
The $\gamma\gamma$ production ratio generally does not deviate much from unity in SUSY.   

Similarly, in the Universal Extra Dimension model, where the $KK$-modes of $W$-boson, 
quarks, and leptons contribute to the loop, at most a 50\% enhancement of $\gamma\gamma$ is found for the 
allowed region of parameters\cite{UED}.

\noindent\underline{\it Possible Dark matter contribution}\ \ \ \ 
When we consider the possible decay to dark matter channel, 
$\Gamma_{h^0}^{\rm tot}$ is replaced by the decay width to the visible channels 
$\Gamma_{h^0}^{\rm vis}$ in LHS of  
Eqs.~(\ref{eq1}), (\ref{eq3}), and (\ref{eq4}), while Eq.~(\ref{eq2}) is unchanged.
The $\Gamma_{h^0}^{\rm tot}$ in Eq.~(\ref{eq2}) now includes the partial 
decay width to the dark matter channel $\Gamma_{h^0\rightarrow D\bar D}$ as\cite{LoganS}
\begin{eqnarray} 
\Gamma_{h^0}^{\rm tot} &=& \Gamma_{h^0}^{\rm vis}+\Gamma_{h^0\rightarrow DD}
\equiv F\cdot\Gamma_{h^0}^{\rm vis},\ \ \ \ 
\Gamma_{h^0}^{\rm vis}\equiv\sum_{A\bar A=b\bar b,\tau^-\tau^+,VV,gg,c\bar c}\Gamma_{h^0\rightarrow A\bar A}\ .
\label{eq6}
\end{eqnarray}
The detection of the Higgs invisible decays at hadron colliders 
has been studied in refs.\cite{Zeppenfeld,Han}. 
The factor $F(\ge 1)$ is related to the $BF$ to dark matter $D$ by 
\begin{eqnarray}
BF(h^0\rightarrow D\bar D) &=& \frac{F-1}{F}.
\label{eq7}
\end{eqnarray}
Our method of fitting the quantities of LHS of  Eq.~(\ref{eq2}) now
determines $\gamma_{AA}^\prime\equiv\gamma_{AA}/F$, not $\gamma_{AA}$, since in Eq.~(\ref{eq2})
$\Gamma_{h^0}^{\rm tot}/\Gamma_{h^0_{\rm SM}}=F\cdot\Gamma_{h^0}^{\rm vis}/\Gamma_{h^0_{\rm SM}}
=F\cdot R$ where $R$ is given by the second equality of Eq.~(\ref{eq3}).

In many models, such as the MSSM in the decoupling limit, the  
$WW^*$ and $ZZ^*$ couplings are nearly the same as those of the SM Higgs boson: $\gamma_{VV}\simeq 1$.\cite{note5} 
In this case, the value of $\gamma_{VV}/F$ obtained by our method 
gives directly the value of $1/F$ which in turn gives $BF(h^0\rightarrow DD)$ following Eq.~(\ref{eq7}).
The best-fit value of $\gamma_{VV}/F$ in Table~\ref{tab3} is $1.34\stackrel{+0.57}{\scriptstyle -0.45}$
which suggests $F\simeq 1$. 
A very large $BF$\cite{Belotsky} to invisible decay channel is 
disfavored\cite{Desai}.
$BF(h^0\rightarrow D\bar D)<0.46$ in 95\% confidence level 
from the present data.


\noindent\underline{\it Concluding remarks}\ \ \ 
We have presented a method of determining the total width of the putative 125~GeV Higgs-boson.
The measurements of the $\gamma\gamma$ cross section of the Higgs signal relative to 
that of the SM will discriminate many candidate models of new physics. 
It is difficult to obtain a theoretical enhancement of the $\gamma\gamma$ signal 
of more than 2. This constitutes a No-Go theorem.
For a charge $Q\le 1$, this theorem is evaded with a new light-mass fermion with color octet(leptogluon)
or color-sextet and a negative Higgs coupling.
Such a colored state can be directly tested\cite{HanLL} by LHC experiments.

Measurements of the vector-boson fusion process and the vector-boson bremsstrahlung processes (c.f. Table~\ref{tab2}) can significantly improve 
the uncertainty on the total Higgs width estimate.

Accurate measurement of the ratio of the $\gamma\gamma$ to $ZZ^*$ cross sections would 
determine $\gamma_{\gamma\gamma}/\gamma_{VV}$, independently of the value of $\gamma_{gg}$.

The branching fraction for the decay of the Higgs boson to dark matter 
can be inferred in the decoupling limit of the $WW^*$ and $ZZ^*$ couplings 
of any two Higgs doublet model\cite{BLS}.

The methods presented in this Letter should be useful when higher statistics data are acquired on the Higgs signal. 
One must be cautious about over-interpreting the data until the Higgs 
 signal is fully established.

\noindent\underline{\it Acknowledgements}

We thank H. Logan and D. Zeppenfeld for drawing our attention to  
Refs.\cite{Zeppen,Belyaev,Duhrssen,Lafaye} 
and for the helpful comments of Patrick Janot.
We benefited from a stimulating discussion with  Yang~Bai.
We thank Zhen Liu for a helpful comment about our results.
M.I. is very grateful to the members of phenomenology institute of University of Wisconsin-Madison for hospitalities.
This work was supported in part by the U.S. Department of Energy under grants No. DE-FG02-95ER40896 and
DE-FG02-12ER41811, 
in part by KAKENHI(2274015, Grant-in-Aid for Young Scientists(B)) and in part by grant
as Special Researcher of Meisei University.

\nocite{*}

\bibliography{apssamp}

\begin{thebibliography}{00}
\bibitem{note1} We could optionally assume one more constraint $\gamma_{bb}=\gamma_{\tau\tau}$
 which is predicted in most Higgs models.
\bibitem{note3} This is not the case when there are many heavy particles. 
 Numerically, $\gamma_{\gamma\gamma}>1$ when an additional 4 generations of quarks and leptons contribute.
\bibitem{note4} In ref.\cite{carena} the light stau effect on the $\gamma\gamma$ cross sections is discussed.
 A 50\% $\gamma\gamma$ enhancement is possible with a very light stau mass $\sim 100$~GeV, 
 close to the LEPII constraint. 
 In  this case, the production rate by the gluon fusion is the same, and 
 the $b\bar b,WW^*,ZZ^*$ cross sections via gluon fusion are almost the same as for the SM Higgs boson.
\end{thebibliography}

\end{document}